\documentclass{mn2e}
\usepackage{psfig}
\usepackage{epsfig}

\newif\ifAMStwofonts

\def\sqiglt{\hbox{\rlap{\lower.55ex \hbox {$\sim$}}\kern-.05em \raise.4ex \hbox{$<$}\,}}
\def\sqiggt{\hbox{\rlap{\lower.55ex \hbox {$\sim$}}\kern-.05em \raise.4ex \hbox{$>$}\,}}
\def\til{\ensuremath{\sim\,}}
\def\w{\ensuremath{\omega}}

\def\chisq{\ensuremath{\chi^2}}
\def\rchisq{\ensuremath{\chi_{\nu}^{2}}}

\newcommand{\tim}[1]{\ensuremath{\times 10^{#1}}}

\def\etal{et al.\ }

\def\mekal{{\sc mekal}}

\def\xmm{\emph{XMM}}
\def\xmmn{\emph{XMM-Newton}}

\def\cps{counts s$^{-1}$}

\title[The intermediate polar PQ~Gem]{\emph{XMM-Newton} observations of the
complex spin pulse of the intermediate polar PQ~Geminorum}

\author[Evans, Hellier \&\ Ramsay]{P.A. Evans$^1$\thanks{pae@astro.keele.ac.uk}, 
Coel Hellier$^1$ and Gavin Ramsay$^2$\\ 
$^1$ Astrophysics Group, School of Chemistry and Physics, Keele
University, Staffordshire, ST5 5BG\\
$^2$ Mullard Space Science Laboratory, University College London,
Holmbury  St.~Mary, Dorking, Surrey RH5 6NT\\
}

\date{Accepted 
      Received }

\pagerange{\pageref{firstpage}--\pageref{lastpage}}
\pubyear{2006}

\begin{document}

\maketitle

\label{firstpage}

\begin{abstract}
The intermediate polar PQ~Geminorum shows a complex pulsation, caused by a
spinning white dwarf, which varies markedly with wavelength. We report \xmmn\/\
observations, including the soft and hard X-ray bands and the first UV
lightcurves of this star.  We update the ephemeris for PQ~Gem allowing us to
align these data with a compilation of lightcurves from the optical to the
X-ray.  Building on work by previous authors, we show how a model in which
accretion flows along skewed field lines, viewed at the correct inclination,
can explain the major features of the lightcurves in all bands.  We discuss how
the skew of the field lines relates to the spinning down of the white-dwarf
rotation.
\end{abstract}


\section{Introduction}
\label{sec:intro}

PQ~Geminorum (RE\,J0751+14) was discovered in the \emph{ROSAT\/} Wide Field
Camera all-sky survey by Mason \etal(1992). They identified it as an
intermediate polar (IP) -- a magnetic subclass of the cataclysmic variable
stars (see, e.g., Patterson 1994; Hellier 2001 for reviews of this class of
star).

The spin-pulse profile of PQ~Gem is complex, and thus presents us with many
clues as to the accretion geometry. It is one of the few IPs to show polarised
light that varies over the spin cycle. It also has a soft blackbody component
to its X-ray emission; out of the twenty-five known IPs, to date only six have
been shown to have such a component (see, e.g., de~Martino \etal2004).  For
this reason it has been much studied, both in optical photometry (Hellier,
Ramseyer \&\ Jablonski 1994), polarimetry (Piirola, Hakala \&\ Coyne 1993; Potter
\etal1997), spectroscopy (Hellier 1997) and X-rays (e.g.\ Duck \etal1994; Mason
1997).

All of these features help us to constrain the geometry of the system. The most
recent model (Potter \etal1997; Mason 1997) suggests that accretion occurs
along magnetic field lines which precede the pole, and that these field lines
pick up material from outside the corotation radius (i.e.\ where the orbital
motion of the disc is slower than the rotation of the white dwarf). Optical
Doppler tomography also shows evidence for accretion along preceding field
lines (Hellier 1997). Accreting from outside the corotation radius is likely to
produce a braking torque on the white dwarf, and this is consistent with the
spin-down reported by Mason (1997). 

We report here an analysis of an \xmmn\/\ observation of PQ~Gem, aimed at testing
the model of Potter \etal(1997) and Mason (1997).

\section{Observations and power spectra}
\label{sec:obsphot}

PQ~Gem was observed for 36 ks by the \xmmn\/\ satellite (Jansen \etal2001)
on 2002 October 7. The EPIC-MOS (Turner \etal2001) and pn (Str\"uder \etal2001)
cameras were operating in Small Window Mode, observing through the Thin Filter.
The Optical Monitor (OM; Mason \etal2001)  observed for 15 ks through the
UVW1 filter  and for 12 ks though the UVM2 filter. The RGS instruments
collected 36 ks of data, however the count-rate was too low to allow for
phase-resolved analysis, so these data are not presented in this paper.

\begin{figure}
\begin{center}
\psfig{file=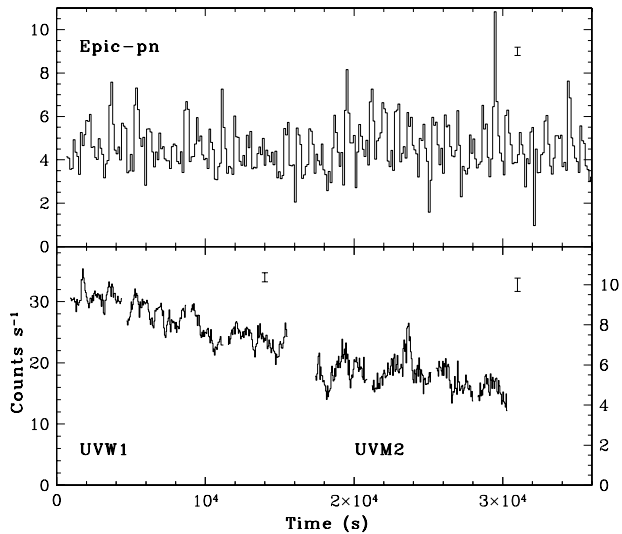,width=8.1cm}
\caption{Top panel: the EPIC-pn X-ray lightcurve binned at 120 s. Lower
panel: The UV lightcurves binned at 50 s (the UVW1 filter has a 
2450--3200-\AA\ bandpass and the UVM2 filter has a 2050--2450 \AA\ bandpass).
The prominent modulation is the 833-s spin period. Typical errors are shown.}
\label{fig:curve}
\end{center}
\end{figure}

We analysed the observation using the {\sc xmm-sas} software v6.0.0, extracting
data from a circular region radius 20 arc sec. Only single- and double-pixel
events were selected. The entire small window of the central CCD in the
EPIC-MOS detectors was affected by source counts, so we used an adjacent chip
to estimate the background. For the EPIC-pn camera, in which only the on-axis
CCD is exposed in Small Window Mode, we used as much area as was free of source
counts to estimate the background.  Some times of high background were
rejected, totalling 850 s in the MOS detectors and 550 in the pn. Lightcurves
are shown in Fig.~\ref{fig:curve}, and are dominated by modulation at the 833-s
white-dwarf spin period. The EPIC-pn camera shows a mean count rate of 4.5
\cps, which implies a 0.2--12 keV flux of 6\tim{-14} W m$^{-2}$. 

\begin{figure}
\begin{center}
\psfig{file=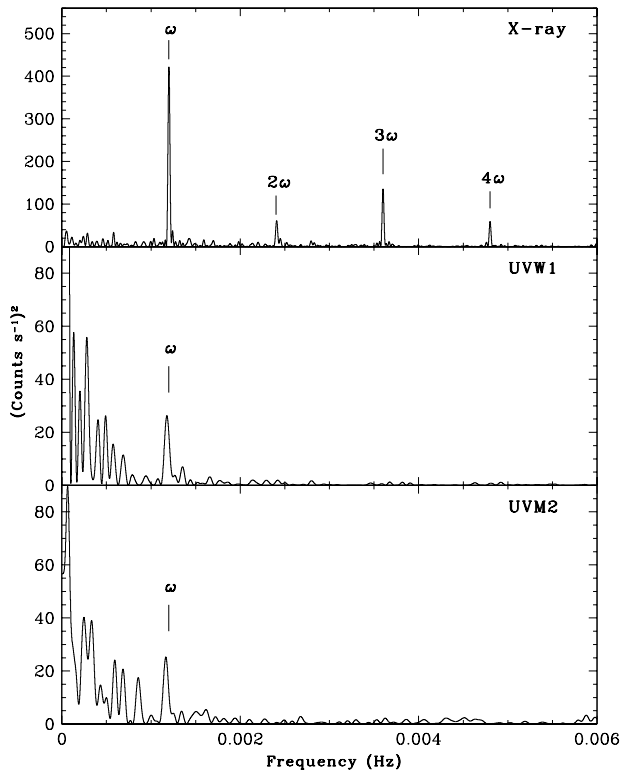,width=8.1cm}
\caption{Power spectra of the X-ray (MOS+pn, upper panel) and  UV (centre and
lower panels) data. The spin frequency is denoted by \w. The low frequency peaks
in the UV data are caused by flickering.} 
\label{fig:ft}
\end{center}
\end{figure}

Fourier transforms (Fig.~\ref{fig:ft}), reveal  spin-period modulation in both
the X-ray and UV emission, but no power is seen at the beat frequency between
the spin and orbital cycles. Power at the beat frequency is seen in the X-ray
power spectra of some IPs and is thought to indicate a portion of the accretion
stream overflowing the disc and coupling to the magnetosphere directly (e.g.\
FO~Aqr, Hellier 1993). However, no previous X-ray observation of PQ~Gem has
revealed power at the beat period, so its absence in our data is not
surprising.

\section{Ephemeris}
\label{sec:ephem}

Mason (1997) showed that the white dwarf in PQ~Gem is spinning down, and
developed a quadratic ephemeris for the spin pulse. By the time of our
observation, this has developed cycle count uncertainties, and thus we update
it using \emph{RXTE}, \emph{ASCA\/} and \emph{Chandra\/} X-ray observations
from the HEASARC archives, and optical observations from the CBA archives
(Kemp, private communication).  For consistency with Mason (1997) we take the
centroid of the soft X-ray dip as the phase zero point. This feature is less
prominent in our data, however from previous work (e.g.\ Mason 1997) it can be
well constrained as occurring 0.2 cycles before soft X-ray maximum. Also,  on
this phasing the hard X-ray dip occurs at phase 0.023, the $B$-band maximum at
0.154, $I$-band maximum at 0.433 and the $V$-band minimum occurs at phase 0.643
(Mason 1997; Hellier \etal1994), allowing us to use all types of data if we
assume that these relative phasings have not changed over the last ten years.
The resultant timings, converted to TDB, are listed in Table~\ref{tab:ephem}. 

\begin{figure}
\begin{center}
\psfig{file=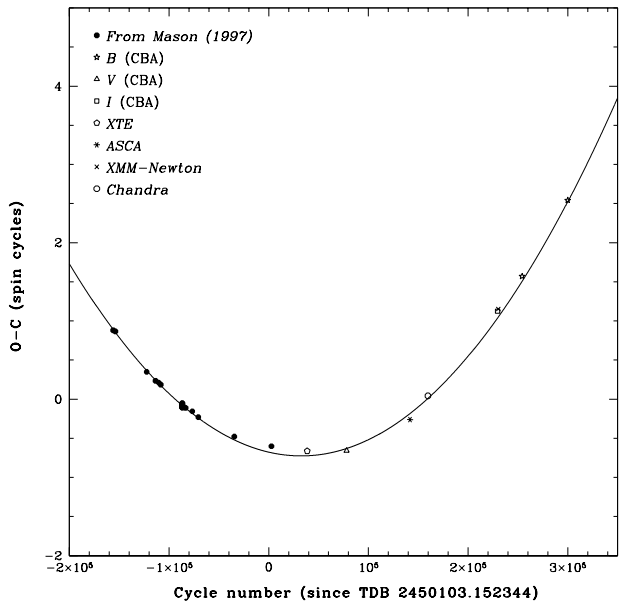,width=8.1cm}
\caption{$O-C$ diagram of PQ~Gem showing the data from Table~\ref{tab:ephem}
and the quadratic ephemeris given in Section~\ref{sec:ephem}.}
\label{fig:ephem}
\end{center}
\end{figure}

The best-fitting quadratic ephemeris calculated is:

\begin{eqnarray*}
{\rm TDB_{dip}} = 245\ 0103.152344(53) &+& 0.0096460639(4) N  \\
 &+& 4.379(26)\tim{-13} N^2 
\end{eqnarray*}

and is shown in Fig.~\ref{fig:ephem}. Adding a cubic term had little effect on
the quality of the fit and is thus not done. Note that this gives a slightly
longer period than Mason (1997), and a smaller value for the period change.

\begin{table}
\begin{center}
\begin{tabular}{cccccc}
\hline
Band                 & Date         &  Duration &  Cycle   &   TDB \\ 
                     &              &  (ks)     &          &   ($-$240\,0000) \\ 
\hline
\emph{RXTE}          & 1997-01-27   &  25.9     &  38620   &  50475.68457 \\
Optical ($V$)        & 1998-02-11   &  6.7      &  78129   &  50856.79208 \\
\emph{ASCA}          & 1999-10-19   &  94.5     &  141826  &  51471.22305 \\
\emph{Chandra}       & 2000-04-09   &  52.3     &  159714  &  51643.77526 \\
Optical ($B$)        & 2002-02-12   &  15.4     &  229678  &  52318.66493 \\
Optical ($I$)        & 2002-02-19   &  22.1     &  230293  &  52324.59754 \\
\xmmn                & 2002-10-07   &  35.7     &  254226  &  52555.46150 \\
Optical ($B$)        & 2003-12-22   &  4.8      &  299997  &  52996.98217 \\
\hline
\end{tabular}
\caption{Timings of phase zero, as defined in the text.}
\label{tab:ephem}
\end{center}
\end{table}

\begin{figure}
\begin{center}
\psfig{file=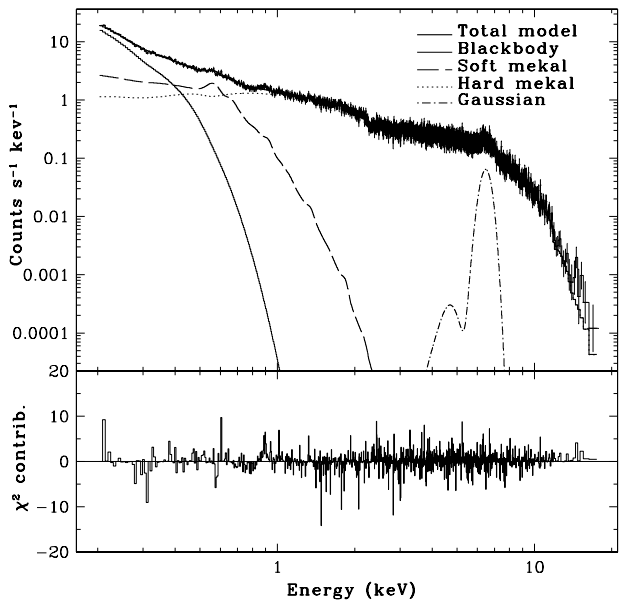,width=8.1cm}
\caption{The phase-averaged spectrum with the fitted model and the
contributions from each component shown.}
\label{fig:comps}
\end{center}
\end{figure}

\section{Spectroscopy}
\label{sec:spec}

We extracted source and background spectra from the regions defined in
Section~\ref{sec:obsphot}, and used the {\sc xmm-sas rmfgen} and {\sc arfgen}
tasks to create the response matrices, for use in {\sc xspec}. When fitting the
spectra we allowed the emission normalisations to optimise independently for
each instrument to minimise the effects of calibration uncertainties between
detectors.

The X-ray emission in an IP arises from plasma heated to X-ray temperatures at
a stand-off accretion shock, which then cools as it approaches the white-dwarf
surface (e.g.\ Aizu 1973; Cropper \etal1999).  We thus initially fitted two
\mekal\ components with different temperatures but an identical abundance,
absorbed by both simple and partial covering absorption. This gave a
best-fitting \chisq\ of 3247 (\rchisq=1.33).  We did not add further \mekal\
components, since the negligible \chisq\ improvement did not justify the
additional free parameters. An alternative model is the full column-fitting
approach of Cropper \etal(1999), however we prefer a simple parameterisation of
the data, allowing us to study spectral changes over the spin pulse with fewer
assumptions about the accretion column structure.

Following previous studies (e.g.\ Mason \etal1992; Duck \etal1994; de~Martino
\etal2004) we added a soft blackbody to our model, improving the fit to a
\chisq\ of 2887 (\rchisq=1.18). Since dense absorption will completely smother
such soft emission, the blackbody was acted on only by a simple absorber,
which was separate from that acting on the \mekal s. 

There were still significant residuals at the 6.4-keV iron fluorescence line,
so we added a narrow Gaussian; this further reduced \chisq\ to 2719
(\rchisq=1.12). The fitted equivalent width ($93.2\pm19$ eV) is comparable with
that found in other IPs observed with \xmm\ (FO~Aqr: 129 eV, Evans \etal2004;
V405~Aur: 121 eV, Evans \&\ Hellier 2004).

The resultant model is shown in Fig.~\ref{fig:comps} and
Table~\ref{tab:average}. Note that some of these parameters may change on the
spin period, and thus the values in the table will be weighted averages.  We
find a best-fitting blackbody temperature of $44\pm9$ eV, in good agreement
with previous studies (Duck \etal1994 found $46^{+12}_{-23}$ eV; de~Martino
\etal2004 reported $56^{+12}_{-14}$ eV). The temperature of the hard \mekal\
component was $39^{+4}_{-6}$ keV, which compares with $17\pm2$ keV from
\emph{BeppoSAX\/} data (de~Martino \etal 2004) and \til70 keV from
\emph{Ginga\/} data (Duck \etal 1994).  The de Martino \etal value is likely to
be most reliable, since the \emph{BeppoSAX\/} detection extends to harder
energies (70 keV versus 12 keV in XMM data), but note that they fitted a
single-temperature model to the entire spectrum, so our results are not
directly comparable.

Our fitted \mekal\ abundance is 0.028 ($\pm 0.004$), which is very low and
reflects the fact that the observed lines are much weaker than expected for the
0.2-keV component included in our model (at the temperature of the hotter
component the plasma is almost entirely ionised, so this component has much
less effect on the abundance than the cooler \mekal). Note that this does
not necessarily indicate a genuinely low metal abundance, but is probably
caused by high opacity in the accretion columns, which will suppress line
emission. de~Martino \etal(2004) found a higher abundance of
$0.33^{+0.07}_{-0.09}$, because they did not include a low-temperature \mekal\
in their model. For comparison with de~Martino et al., we tried removing the
softer \mekal\ component from our model, however this caused \chisq\ to
increase by \til1200.

{\sc xspec} also provides a {\sc cemekl} emission model, which reproduces
\mekal\ emission along a continuous distribution of temperatures, with
emissivity varying according to $(T/T_{\rm max})^\alpha$. Exchanging the two
single-temperature \mekal s for a {\sc cemekl} caused a very large (\til800)
increase in \chisq, suggesting that emissivity in the accretion column does not
have a power-law dependence on temperature.

\begin{table}
\begin{center}
\begin{tabular}{lccc}
\hline
Component &  Parameter                & Value               & Error \\ 
          &  (Units) \\
\hline
Absn.     &  $n_{\rm H}$ (cm$^{-2}$)  & 4.75\tim{19}        & (+1.74, $-$2.02) \\ 
Blackbody &  $kT$ (eV)                & 43.9                & (+8.7, $-$8.2)   \\ 
          &  Norm (pn)                & 2.44\tim{-4}        & (+0.34, $-$0.35) \\
Absn.     &  $n_{\rm H}$ (cm$^{-2}$)  & 9.67\tim{20}        & (+1.26, $-$1.95) \\
Partial   &  $n_{\rm H}$ (cm$^{-2}$)  & 11.1\tim{22}        & (+0.88, $-$0.73) \\
\ \ \ Absn.  &  CvrFract              & 0.450               & (+0.006, $-$0.008) \\
Gaussian  &  Energy (keV)             & 6.40                & frozen           \\
          &  Norm (pn)                & 2.43\tim{-5}        & (+0.42, $-$0.41) \\
          &  Eq. Wid (pn, eV)         & 93.2                & (+17.8 $-$17.5) \\
Mekal     &  $kT$ (keV)               & 0.179               & ($\pm$0.002) \\
          &  Abundance                & 2.82\tim{-2}        & (+0.58, $-$0.20) \\
          &  Norm (pn)                & 5.84\tim{-2}        & (+1.45, $-$1.29) \\
Mekal     &  $kT$ (keV)               & 39.4                & (+4.1, $-$6.3) \\
          &  Norm (pn)                & 1.77\tim{-2}        & (+0.04, $-$0.03) \\

\hline
\end{tabular}
\caption{Model components and parameters fitted to the phase-averaged spectrum
(see Section~\ref{sec:spec}). The errors, which are quoted to the same power of
ten as the corresponding value are the 90\% confidence errors according to
formal statistics; note however that these are likely underestimates, since
they do not account for the calibration systematics.}
\label{tab:average}
\end{center}
\end{table}

\section{Phase-resolved spectroscopy}
\label{sec:spin}

\begin{figure}
\begin{center}
\psfig{file=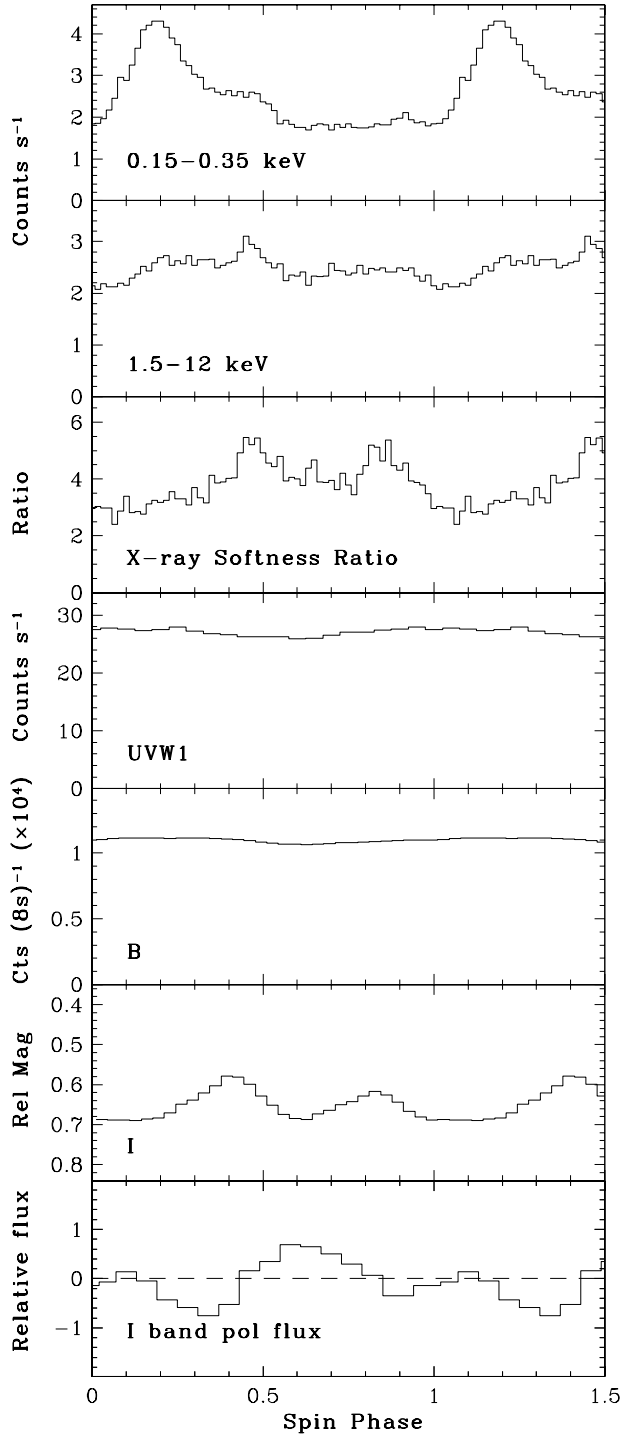,width=8.1cm}
\caption{Spin folds of PQ~Gem. The upper two panels show the soft and hard
X-ray folds. The softness ratio is defined as (1.5--4)/(6--12) keV. The $B$ and
$I$ band data are taken from Hellier \etal(1994), and the polarised flux is
found from the $I$ band flux in Hellier \etal(1994) and the polarimetry of
Potter \etal(1997). Phase zero corresponds to the X-ray dip, as in our
ephemeris (Section~\ref{sec:ephem}).}
\label{fig:spin}
\end{center}
\end{figure}

We present spin-folded light-curves and a softness ratio in
Fig.~\ref{fig:spin}. In previously reported \emph{ROSAT\/} data (Mason 1997),
the soft X-rays show a dip at phase 0, a broad maximum centred on phase 0.2 and
a `shoulder' around phases 0.4--0.5. The harder X-rays show the dip and also a
`spike' coincident with the soft shoulder. Our data are similar, except that
the X-ray dip is shallower. This was most prominent in \emph{ROSAT\/} data
(from 1993; Mason 1997), less so in \emph{BeppoSAX\/} data (from 1996;
de~Martino \etal2004), and still less prominent in our \xmm\/\ data (from
2002).

To investigate these spectral changes we applied two models to four phase
regions (the dip, the spike, and phases 0.2--0.4 and 0.6--0.9), testing whether
the changes are reproduced by variations in absorption, model normalisation, or
both. The first model was that developed in Section~4, in which the \mekal\ 
temperatures and normalisations can vary with respect to one-another. The second
model was the stratified column model of Cropper \etal(1999), which uses
multiple \mekal s with temperatures and relative normalisations determined by
the physics of an accretion column. 

Both models gave adequate fits if both their normalisations and the absorption
are allowed to vary on the spin cycle. An example fit is given in
Table~\ref{tab:pres}. Neither model gave an acceptable fit if only the
absorption was allowed to change (with $\Delta\chisq$\sqiggt\,40 when this was
attempted, compared to a fit where all parameters can vary). We then tried
instead forcing the absorption to remain constant across the spin cycle. The
stratified column model did not give an acceptable fit in this case, however
the greater freedom of the unconstrained \mekal\ model did allow acceptable
fits (\chisq\ poorer by no more than 20) except during the dip. If we accept
that the relative \mekal\ normalisations must be constrained as in the
stratified column model, this implies that the absorption must be allowed to
vary between phase regions.

\begin{table*}
\begin{center}
\begin{tabular}{lcccc}
\hline
Parameter (units)                        &  Dip                     &                        & Spike   \\
\ \ \ Phase region                       & 0.95--1.05               & 0.2--0.4               & 0.4--0.5               & 0.6--0.9 \\
\hline 
Abs $n_{\rm H}$ (10$^{19}$ cm$^{-2}$)    & 16.0 (+23.6, -8.01)      & 2.4 (+6.2, -2.4)       & 7.1 (+14.1, -7.1)      & 13.3 (+13.1, -5.2) \\
\\
B-body  $kT$ (eV)                        & 37.4 (+1.9, -7.4)        & 45.9 (+3.2, -3.4)      & 41.5 (+2.5, -6.6)      & 39.8 (+5.2, -5.6)
\\
           Norm (pn \tim{-4})            & 4.2 (+21.7, -2.32)       & 3.1 (+1.7, -2.5)       & 3.0 (+3.3, -1.3)       & 2.8 (+3.5, -1.4) \\
\\
Abs $n_{\rm H}$ (10$^{21}$ cm$^{-2}$)    & 0.86 (+0.29, -0.19)      & 1.16 (+0.20, -0.26)    & 0.91 (+0.22, -0.30)    & 1.05 (+0.24, -0.29) \\
\\  
PCF $n_{\rm H}$ (10$^{22}$ cm$^{-2}$)    & 8.2 (+2.1, -1.7)         & 12.7 (+2.2, -2.0)      & 9.4 (+3.2, -2.2)       & 12.7 (+2.5, -2.2)
\\
\ \ \ CvrFract                           & 0.526 (+0.032, -0.030)   & 0.529 (+0.032, -0.026) & 0.343 (+0.044, -0.021) & 0.393 (+0.039, -0.031)
\\
\\
Mekal $kT$ (eV)                          & \multicolumn{4}{c}{177$\pm3$, tied across all regions} \\
\ \ \  Abundance (\tim{-2})              & \multicolumn{4}{c}{2.40 (+0.53, -0.49), tied across all regions} \\
\ \ \  Norm (pn \tim{-2})                & 6.74 (+0.37, -0.34)      & 8.70 (+0.34, -0.38)    & 7.23 (+0.36, -0.34)    & 6.83 (+0.26, -0.25) \\
\\
Mekal  $kT$ (keV)                        & \multicolumn{4}{c}{38.7 (+8.5, -8.0), tied across all regions} \\
\ \ \  Norm (pn \tim{-2})                & 1.615 (+0.088, -0.0082)  & 2.083 (+0.081, -0.092) & 1.731 (+0.087, -0.081) & 1.634 (+0.061, -0.059) \\
\\
\hline
\end{tabular}
\caption{The model parameters when the four phase regions are fitted
simultaneously, with absorption and normalisation free to vary. The \mekal\
temperatures and abundance were tied between regions, as was the ratio of their
emission normalisations.}
\label{tab:pres}
\end{center}
\end{table*}

\section{Discussion}
\label{sec:disc}

Potter \etal(1997) and Mason (1997) developed a model for PQ~Gem based on
optical photometry, polarimetry and X-ray observations. In this model the
accreting field lines at the upper magnetic pole lie in our line of sight at
phase zero, obscuring our view of the X-ray emitting accretion footprints and
causing the X-ray dip. However, to explain asymmetries in the lightcurve, the
accretion needs to be predominantly along field lines preceding the pole; thus
the magnetic pole itself is not on the white-dwarf meridian until phase 0.1. 

\begin{figure*}
\begin{center}
\psfig{file=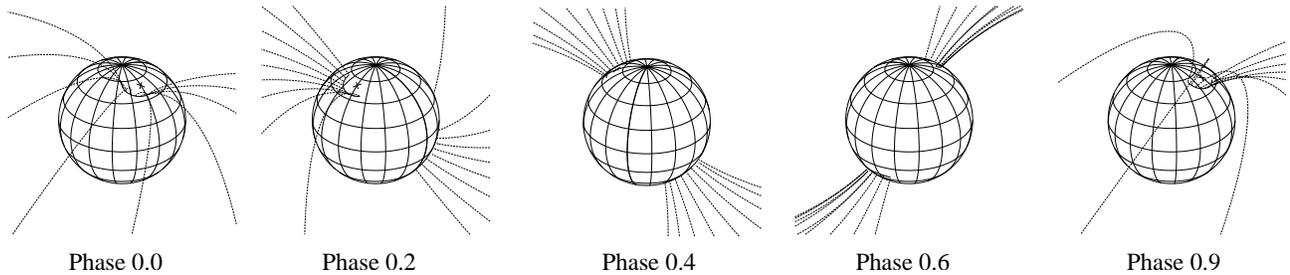,width=17cm}
\caption{Schematic diagrams during the X-ray dip (phase 0.0), blackbody maximum
(phase 0.2), the `spike' (phase 0.4), phase 0.6, and phase 0.9.}
\label{fig:schem}
\end{center}
\end{figure*}

By phase 0.1 the curtains are moving out of our line of sight, so the X-ray and
UV flux start to rise (Fig.~\ref{fig:schem}). By phase 0.2 the accretion
curtains have moved on, giving us a relatively unobscured view of the heated
white-dwarf surface near the accretion column.  This results in soft-X-ray
maximum.

While the soft X-ray emission comes from the heated white-dwarf surface, the
$I$-band flux comes from cyclotron emission in the accretion columns above
the surface. We see maximum $I$-band flux  when the magnetic poles are across
our line of sight, and we can see emission from both poles simultaneously
(e.g.\ phase 0.4, see Figs.~\ref{fig:spin} and \ref{fig:schem}). $I$-band
minima occur when the magnetic axis points towards us and we see only one
pole.  

\subsection{Implications of our data} \label{sec:discxmm}

In Section~5 we concluded that the absorption varies over the spin cycle. 
This can be explained by the accretion curtain model, in which
opacity along the line of sight is highest when the curtain points towards us
(phase 0 in Fig 6) and lowest half a cycle later (phase 0.5), 
producing a sinusoidal variation in the  softness ratio
(see, e.g., Hellier, Cropper \&\ Mason 1991, and the modelling by
Kim \& Beuermann 1995). 

The overall softness variation in Fig.~\ref{fig:spin} has the correct phasing
to be explained as above, however there is a reduction in softness during
phases 0.5--0.8, giving apparent maxima at phases 0.45 and 0.85.  We suggest
that this could arise if the base of the upper accretion column has passed over
the white-dwarf limb (see Fig.~\ref{fig:schem}) such that we don't see the
softest emission from the cooler regions near the base of the column. To
complete the explanation we would also need an asymmetry between the upper and
lower poles so that the base of the lower pole did not appear and produce an
opposite effect; alternatively, if there were considerable opacity in the
orbital plane, it could prevent soft emission at the lower pole being seen.

In the hard X-ray pulse profile we see a `spike' at phases 0.4--0.5, coincident
with a `shoulder' in the soft X-ray decline. Mason (1997) suggested that the
lower accretion region could be larger than the upper one, thus at phases
0.4--0.5 we could be seeing both columns and footprints, explaining these
features.    We also note that the  softness rises during phases 0.2--0.4,
while the flux remains approximately constant, for which we do not have an
explanation. 

Turning to other wavelengths: the UV emission could arise from both the heated
white-dwarf surface and the X-ray irradiated accretion curtains. In principle,
more flux could be visible when the upper footprint points towards us, however
this is countered by obscuration of this region by the accretion curtains (see
Fig.~\ref{fig:schem}), and the combination of these effects could explain the
relatively low pulse fraction of \til7 per cent (Fig.~\ref{fig:spin}).

\begin{figure}
\begin{center}
\psfig{file=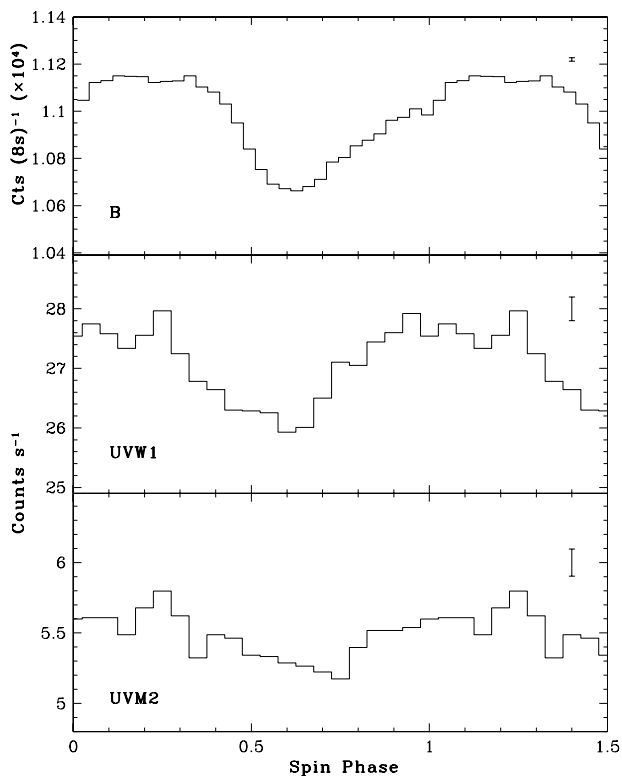,width=8.1cm}
\caption{The pulse-profile of PQ~Gem in the UV and $B$ bands. The scale here
has been enlarged to show the shape of the pulses. Typical photon-noise errors
are shown, though uncertainties owing to flickering are likely
to be larger (see Fig.~2).}
\label{fig:spin2}
\end{center}
\end{figure}

However, since we do see more flux at phases 0.9--0.3 (Fig.~\ref{fig:spin2}),
it is possible that some of the upper heated polecap is visible above the sweep
of the curtain (see, e.g., the phase 0 panel of Fig.~\ref{fig:schem}), as we
have already argued to explain the soft X-ray lightcurve.

The pulsed $B$-band light can arise from a combination of the heated white
dwarf, the accretion curtains, and the irradiated disc.  The pulse has a depth
of only \til4 per cent (Fig.~\ref{fig:spin}), which suggests these components
act in anti-phase. 

\subsection{Field-disc interaction}

The fact that the white dwarf in PQ~Gem is spinning down (Section~3) implies
that it is not in equilibrium. Further, Mason (1997) suggests that the
accreting field lines thread the accretion disc a little outside the corotation
radius, so that the slower-moving disc material exerts a spin-down torque on
the dipole. We would then expect the field lines to be swept backwards,
particularly near the threading region where magnetic and material pressures
are roughly balanced. A discussed by Mason (1997), it is hard to see how a
field line that is swept back and moving faster than the local disc can scoop
up material. However, a field line ahead of the pole will meet the disc at an
inclined angle that counteracts any effect of being swept back. Thus accreting
along leading field lines may be a consequence of the fact that the white dwarf
is  spinning faster than equilibrium. 

Testing such ideas by applying them to other IPs is hampered by the fact that
in few systems do the spin-pulse lightcurves give sufficient observational
clues for us to deduce the layout of the accreting field lines. However, in
FO~Aqr we found that the accretion is predominantly along field lines swept
back from the magnetic pole (e.g.\ Evans \etal2004). This lag, however, appears
to persist both when the white dwarf  is spinning up and when it is spinning
down, which  argues against any simplistic interpretation.   Thus, study of 
intermediate polars is giving us observational clues to the  difficult topic of
how a magnetic field interacts with an  accretion disc, but study of a greater
sample of systems is needed to make further progress. 

\section*{Acknowledgements}

This research has made use of data obtained through the High Energy
Astrophysics Science Archive Research Center Online Service, provided by the
NASA/Goddard Space Flight Center. We also thank Jonathan Kemp and Joe Patterson
for providing us with optical photometry of PQ~Gem from the CBA archives. The
\xmmn\/\ observations were obtained as part of the Optical Monitor Guaranteed
Time programme.

\label{lastpage}
\end{document}